\documentclass[a4paper, amsfonts, amssymb, amsmath, reprint, showkeys, nofootinbib, twoside,notitlepage,onecolumn]{revtex4-1}

\usepackage{amsmath,amstext}
\usepackage[T1]{fontenc}
\usepackage{amssymb}
\usepackage{graphicx}
\usepackage{ae,aecompl}

\DeclareFontFamily{OT1}{pzc}{}
\DeclareFontShape{OT1}{pzc}{m}{it}{<-> s * [1.10] pzcmi7t}{}
\DeclareMathAlphabet{\mathpzc}{OT1}{pzc}{m}{it}

\usepackage{hyperref}
\usepackage{amsmath}
\usepackage{amssymb}
\usepackage{mathtools}
\usepackage{bm}
\usepackage{cleveref}
\usepackage{tensor}
\usepackage{braket}
\usepackage{enumitem}
\usepackage{mhchem}
\usepackage{amsthm}
\usepackage{nccmath}
\usepackage{mathrsfs}
\usepackage{color}

\def\be{\begin{equation}}
\def\ee{\end{equation}}
\def\beq{\begin{eqnarray}}
\def\eeq{\end{eqnarray}}

\theoremstyle{definition}

\theoremstyle{theorem}

\theoremstyle{corollary}

\begin{document}
\title{How acausal equations emerge from causal dynamics}
\author{L.~Gavassino}
\affiliation{Department of Applied Mathematics and Theoretical Physics, University of Cambridge, Wilberforce Road, Cambridge CB3 0WA, United Kingdom}

\begin{abstract}
We construct a causal and covariantly stable kinetic model whose spectrum at real wavenumbers $k$ reproduces any rest-frame stable dissipative dispersion relation $\omega(k)$ via suitable initialization of the microscopic degrees of freedom. Macroscopic observables can therefore obey arbitrary linear evolution equations (including forms that would be acausal if taken as fundamental), while the underlying dynamics remains causal, and all apparent propagation is encoded in the initial data. This provides an explicit counterexample to the idea that microscopic causality alone constrains the analytic form of dispersion relations at real $k$. In particular, bounds on transport coefficients based solely on the analytic structure of $\omega(k)$, such as the hydrohedron bounds, require additional assumptions about the region in the complex $k$-plane where $\omega(k)$ corresponds to physical modes.
\end{abstract} 
\maketitle

\section{Introduction}
\vspace{-0.3cm}

One of the oldest questions in relativistic many-body physics is whether the principle of causality (i.e. the absence of superluminal information transfer) imposes universal constraints on the dispersion relations of a linearized theory. 
Early attempts to bound phase, group or front velocities have all failed, as counterexamples were later identified \cite{Sommerfeld1907,Brillouin1960,BludmanRuderman1968,Susskind1969,Fox1970,Krotscheck1978,Pu2010,GavassinoDisperisons2023mad}.

The underlying difficulty is that, in systems with multiple degrees of freedom, focusing on a \textit{single} dispersion relation $\omega(k)$ can be misleading. The apparent superluminal propagation of a wavepacket may either reflect genuine signal transmission, or simply arise from a coordinated evolution of microscopic degrees of freedom that were initialized to mimic propagation while evolving locally \cite{GavassinoDisperisons2023mad}. The standard analogy is the stadium wave: a pattern can move at speed $2c$ if spectators at position $x$ agree \textit{in advance} to stand up at time $t\,{=}\,x/(2c)$, with no real signal being propagated.

\begin{figure}[b!]
    \centering
\includegraphics[width=0.42\linewidth]{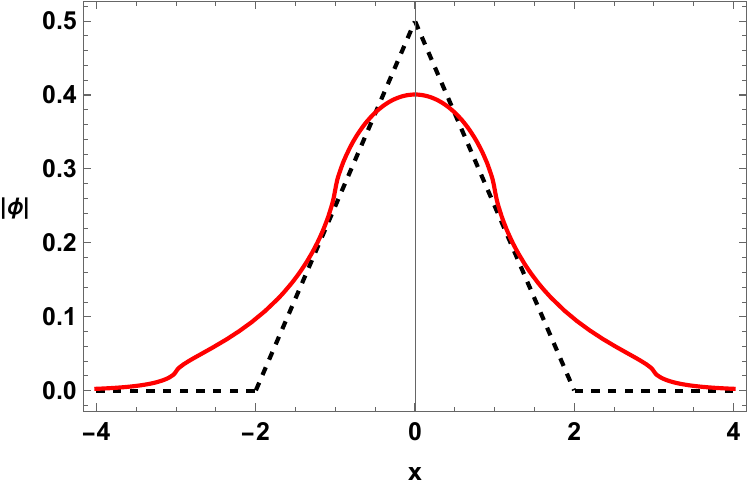}\hspace{0.08\linewidth}
\includegraphics[width=0.42\linewidth]{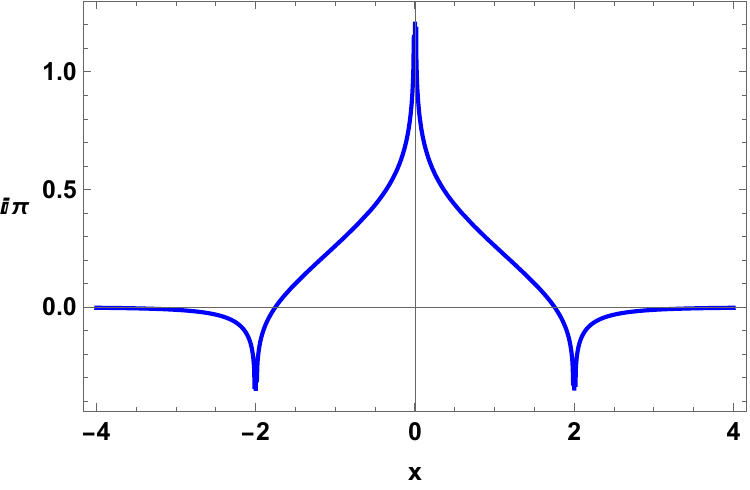}
\caption{Apparent superluminality in the Klein--Gordon equation. \textit{Left panel}: The wavepacket \eqref{wavepacket} is confined to the interval $[-2,2]$ at $t=0$ (black, dashed), but at $t=1$ it has spread beyond $[-3,3]$ (red), giving the impression of superluminal propagation. \textit{Right panel}: The paradox is resolved by noting that the conjugate field $\pi=\partial_t\phi$ is already nonzero outside $[-2,2]$ at $t=0$. Thus, the ``superluminal'' tails of $\phi$ do not carry new information, but simply reflect the nonlocal initialization of $\pi$.}
    \label{fig:NotReallyAcausal}
\end{figure}

This kind of phenomenon can also arise in explicit calculations. Consider the wavepacket
\vspace{-0.1cm}
\begin{equation}\label{wavepacket}
\phi(t,x)=\int_{\mathbb{R}} \dfrac{dk}{2\pi} \,\text{sinc}(k)^2\, e^{ikx-i\sqrt{1+k^2}\, t} \, ,
\end{equation}
which is an exact solution of the Klein--Gordon equation $(\partial_x^2-\partial_t^2)\phi=\phi$, a manifestly causal theory. At $t=0$, the function $\phi$ is supported on the interval $[-2,2]$. Yet, for any $t>0$, one finds that its support extends over the entire real line (see Fig.~\ref{fig:NotReallyAcausal}, left panel), giving the impression of superluminal spreading \cite{Hegerfeldt1974}.
This, however, does not violate causality as the Klein--Gordon equation is of second order in time, and its initial data consist of both $\phi$ and $\pi=\partial_t\phi$. Computing the latter, one finds that $\pi(0,x)$ has unbounded support from the outset\footnote{\textit{Formal proof} \cite{GavassinoDisperisons2023mad}: Since the Fourier transform of a compactly supported function is necessarily entire, $\phi(t,x)$ cannot remain compactly supported at any $t>0$, due to the square root $\sqrt{1+k^2}$ in the exponential. The same non-analyticity also implies that $\pi(0,x)=\partial_t \phi(0,x)$ is not compactly supported, as the time derivative introduces a factor $\sqrt{1+k^2}$ in Fourier space.} (see Fig.~\ref{fig:NotReallyAcausal}, right panel). The apparent superluminal expansion of \eqref{wavepacket} is therefore entirely due to the nonlocal initialization of $\pi$.

Recently, the authors of \cite{HellerBounds2022ejw,HellerHydrohedron2023jtd} proposed a novel and ingenious approach to place bounds on dispersion relations. They observed that if a pair $(\omega,k)\in\mathbb{C}^2$ is an excitation mode of a causal (and stable \cite{GavassinoBounds2023myj,HoultClassicalDispersion:2023clg,Wang:2023csj}) theory, then it satisfies
\begin{equation}\label{StabilityBound}
\mathfrak{Im}\,\omega \leq |\mathfrak{Im}\,k| \, .
\end{equation}
They then assumed that if a dispersion relation $\omega(k):\mathbb{R}\to\mathbb{R}$ admits a series expansion around $k=0$, $\omega=\sum_n c_n k^n$, with radius of convergence $\mathcal{R}$, it should be possible to analytically continue it to complex $k$ within $|k|<\mathcal{R}$. Applying \eqref{StabilityBound} to this continuation, they derived bounds on the coefficients $c_n$ in terms of $\mathcal{R}$ (the ``hydrohedron bounds'' \cite{HellerHydrohedron2023jtd}). For instance, for a diffusive mode $\omega=-i\mathfrak{D}k^2+\dots$, they obtained the bound
\(
\mathfrak{D} \mathcal{R}\leq \frac{16}{3\pi} \, .
\)
Unfortunately, this inequality has also been recently shown not to be universally valid. In \cite{GavassinoDiffusionCompatible:2026tvy,GavassinoFokkerPLanck:2026zsz}, it was shown that the hydrodynamic mode of relativistic Fokker--Planck kinetic theory (a causal and stable model) is exactly $\omega=-i\mathfrak{D}k^2$, for all real $k$. In this case, $\mathcal{R}=+\infty$, but $\mathfrak{D}$ remains finite (see \cite{Brants2025SavingCausality} for a similar result). The reason the argument of \cite{HellerBounds2022ejw,HellerHydrohedron2023jtd} can fail (in its original formulation) is that, although a dispersion relation can be analytically continued into the complex plane, the continuation may not correspond to genuine quasi-normal modes of the microscopic theory. In the relativistic Fokker--Planck case, the hydrodynamic dispersion $\omega=-i\mathfrak{D}k^2$ is an actual mode within the strip $|\mathfrak{Im}\,k|\leq 1/(2\mathfrak{D})$ \cite{GavassinoFokkerPLanck:2026zsz}, and therefore ceases to exist before entering the region where \eqref{StabilityBound} would be violated.

In this work, we show that the Fokker-Planck counterexample to the hydrohedron bounds is just one particular instance of a more general fact. Namely that, if one restricts attention only to real $k$, then it is possible to embed any rest-frame stable dissipative fluid model (even if acausal) into a larger causal and stable kinetic theory. In particular, we can construct a ``stadium-wave'' kinetic model, in which neighboring points do not exchange information, yet each point possesses sufficiently many relaxing and oscillatory degrees of freedom to reproduce any rest-frame stable dispersion relation $\omega(k)$ at real $k$ through an appropriate choice of initial data.

In the following, we work in $(1{+}1)$-dimensional Minkowski spacetime and adopt natural units with $c=1$.

\vspace{-0.4cm}
\section{A stadium-wave model}\label{stadiumwave}
\vspace{-0.3cm}

We consider a kinetic toy model involving a linearized observable $\psi(t,x,\varepsilon)$, which counts the number of particles at the spacetime point $(t,x)$ and having energy $\varepsilon\in[0,+\infty)$. Then, we postulate the equation of motion
\begin{equation}\label{model}
\partial_t \psi = \partial_\varepsilon \psi \, ,
\end{equation}
whose general solution is
\vspace{-0.2cm}
\begin{equation}\label{psiPt}
\psi(t,x,\varepsilon)=\psi(0,x,\varepsilon+t) \, ,
\end{equation}
which describes a process in which particles remain at a fixed spatial position while progressively losing energy, and eventually disappearing once their energy reaches $0$.

Equation \eqref{model} is manifestly causal: its spacetime characteristics are the lines $x=\mathrm{const}$, so no information propagates between different spatial points (see \cite{DudynskiCausality1985,Bemfica2019_conformal1,GavassinoSuperlum2021,DisconziReview:2023rtt} for the definition of causality in partial differential equations). In this sense, the particles behave like the spectators in the stadium: they do not communicate, but evolve locally based solely on their initial information.
To establish covariant stability and dissipation, it suffices to exhibit a quadratic timelike future-directed current with non-positive divergence, whose flux across Cauchy surface plays the role of a Lyapunov function \cite{Hishcock1983,GavassinoLyapunov_2020,GavassinoGibbs2021,GavassinoCausality2021,lasalle1961stability}. In this case, one may take
\begin{equation}
E^\mu = \delta^\mu_t \int_0^\infty |\psi|^2 d\varepsilon \qquad \qquad \Longrightarrow \qquad 
\qquad 
\partial_\mu E^\mu = -|\psi|^2_{\varepsilon=0} \leq 0 \, .
\end{equation}

The key feature that makes the above theory interesting is its quasi-normal spectrum. Within the Hilbert space $\psi(\varepsilon)\,{\in}\, L^2[0,+\infty)$ (i.e. within the space of states with finite $E^0$), the functions
\begin{equation}\label{magic}
\psi^{\{k,\omega\}}(t,x,\varepsilon)
= e^{ikx-i\omega(\varepsilon+t)} \, ,
\qquad
k\in\mathbb{C}, 
\quad
\omega\in\mathbb{R}-i\mathbb{R}_+ \, ,
\end{equation}
are quasi-normal modes. Thus, for every $k$, the excitation spectrum fills the entire lower half-plane $\mathfrak{Im}\,\omega<0$, which lies within the causality-stability region \eqref{StabilityBound}. Crucially, the modes \eqref{magic} are normalizable in $L^2[0,+\infty)$ (i.e. $E^0<\infty$), and all their energy moments $\int_0^\infty \varepsilon^n \psi d\varepsilon$ ($n\,{\in}\, \mathbb{N}$) are finite. Thus, they represent genuine physical states \cite{Groot1980RelativisticKT}, and belong to the point spectrum (i.e. are proper eigenvalues in the relevant Hilbert space). Note that the existence of~\eqref{magic} also implies that $\partial_\varepsilon{+}i\omega$ is not invertible for $\mathfrak{Im}\omega\,{<}\,0$, so \eqref{magic} also correspond to correlator singularities (see App. \ref{AppAAA}).

This structure allows one to ``carve out'' arbitrary (rest-frame stable) dispersion relations by appropriate choices of initial data (i.e. by coordinating a stadium wave). In particular, given any function $f:\mathbb{R}\to\mathbb{R}-i\mathbb{R}_+$, any convergent Fourier superposition of the modes $\psi^{\{k,f(k)\}}$ solves \eqref{model}, and is governed by the dispersion relation $\omega=f(k)$. 
Moreover, the total particle density $\rho=\int_0^\infty \psi d\varepsilon$ (and likewise any slice of $\psi$ at fixed $\varepsilon$) satisfies the evolution equation
\begin{equation}
i\partial_t \rho = f(-i\partial_x)\,\rho \, ,
\end{equation}
which is in general non-local and acausal. However, no information is transmitted: each spatial point evolves independently, and the apparent propagation is entirely encoded in the initial data. This is simply a stadium wave.

\subsection{An explicit example}
\vspace{-0.3cm}

\begin{figure}[b!]
    \centering
\includegraphics[width=0.43\linewidth]{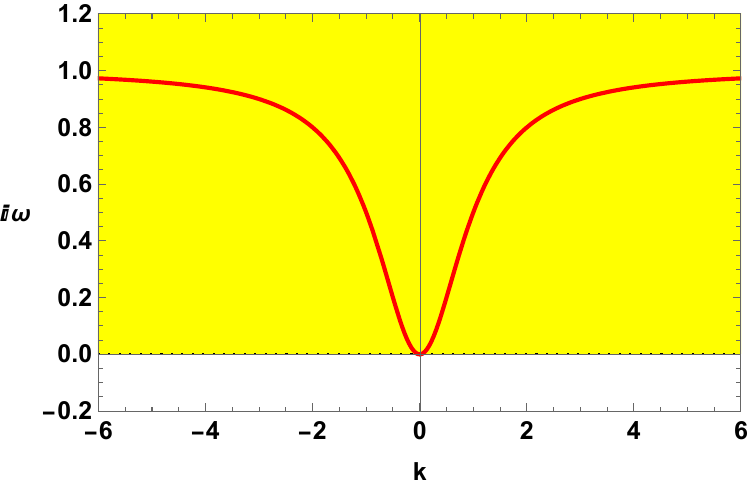}\hspace{0.08\linewidth}
\includegraphics[width=0.45\linewidth]{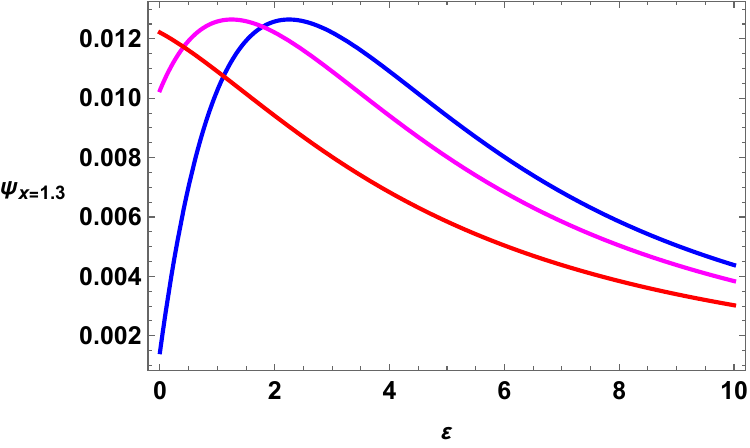}
\includegraphics[width=0.43\linewidth]{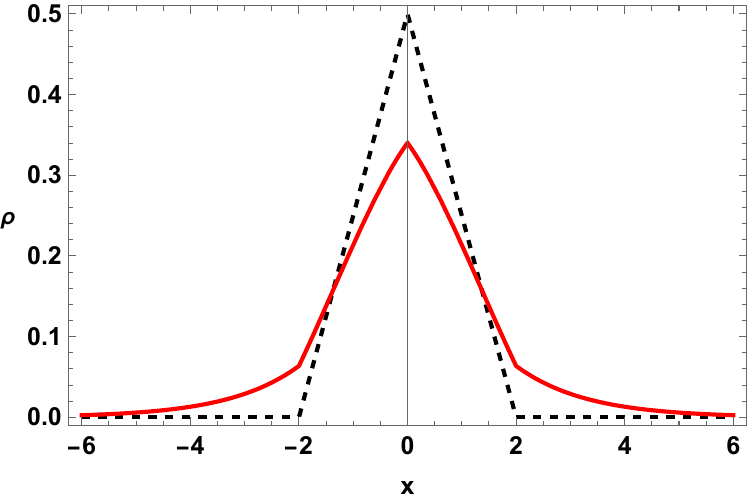}\hspace{0.08\linewidth}
\includegraphics[width=0.45\linewidth]{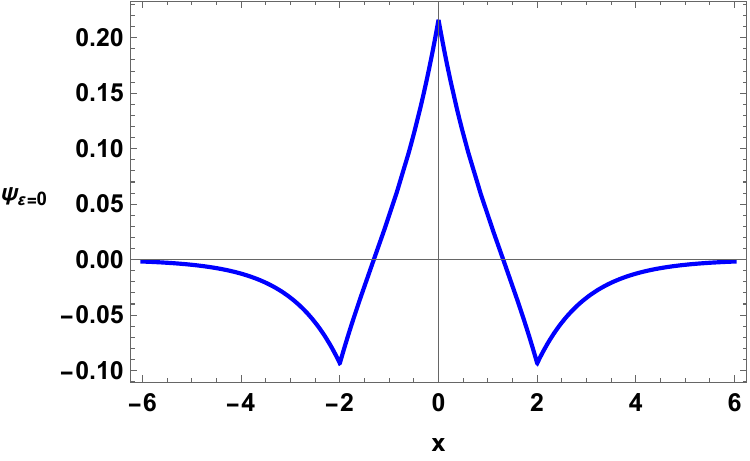}
\caption{How to fake superluminal spread using model \eqref{model}. \textit{Top left}: The spectrum of the theory contains all relaxation rates from $0$ to $+\infty$ (yellow region) at all $k$. One can therefore construct superpositions corresponding to any dissipative dispersion relation, e.g. $\omega=-ik^2/(1+k^2)$ (red line). \textit{Top right}: The solution \eqref{Stadiumpacket} evolves purely locally, with particles at each spatial point $x$ losing energy according to \eqref{psiPt} and disappearing past $\varepsilon=0$. Here we show the evolution at $x=1.3$ for $t=0$ (blue), $1$ (magenta), and $3$ (red). \textit{Bottom left}: The initial data are arranged so that the total density reproduces the propagation of a triangular profile (dashed) according to \eqref{nonlocal}, a non-local equation. The red curve (at $t=1$) exhibits long tails despite the initial data being supported in $[-2,2]$. \textit{Bottom right}: As in Fig.~\ref{fig:NotReallyAcausal}, the apparent superluminal propagation originates from degrees of freedom that are already non-zero outside $[-2,2]$ at $t=0$ (here, the $\psi$-slice at $\varepsilon=0$). Note that the negativity of $\psi$ is immaterial, as the linearity of the theory allows for an arbitrary constant shift.}
    \label{fig:FakingSuperlum}
\end{figure}

Let us consider an explicit example. Take the momentum-suppressed diffusion equation
\(
(1-\partial_x^2)\partial_t \rho=\partial_x^2\rho
\) \cite{Barenblatt1960BasicCI,Barenblatt1963OnCB,Aifantis1980}.
The lines $x=\mathrm{const}$ are characteristics of this equation, which is therefore acausal, if viewed as a stand-alone theory. Indeed, it is highly non-local: for functions bounded at spatial infinity, it can be equivalently rewritten as \cite{Ting197423}
\vspace{-0.1cm}
\begin{equation}\label{nonlocal}
\partial_t \rho =(1-\partial_x^2)^{-1}\partial_x^2 \rho
= \int_{\mathbb{R}} \dfrac{e^{-|\xi|}}{2} \,\partial_x^2 \rho(x{+}\xi)\, d\xi \, .
\end{equation}
However, its dispersion relation $\omega=-ik^2/(1{+}k^2)$ is stable in the frame $(t,x)$, and can therefore be reproduced within the model \eqref{model}. Indeed, integrating \eqref{magic} over all $\varepsilon\geq 0$ produces a factor $(i\omega)^{-1}$, so any Fourier-decomposable solution of \eqref{nonlocal} is the density of a corresponding superposition of \eqref{magic} with a compensating factor $\frac{k^2}{1+k^2}$. For instance, the solution
\vspace{-0.1cm}
\begin{equation}\label{Stadiumpacket}
\psi(t,x,\varepsilon)=\int_{\mathbb{R}} \dfrac{dk}{2\pi} \,\text{sinc}(k)^2 \dfrac{k^2}{1+k^2}\, e^{ikx-\frac{k^2}{1+k^2} (\varepsilon+t)} \, \qquad \Longrightarrow \, \qquad \rho(t,x)=\int_{\mathbb{R}} \dfrac{dk}{2\pi} \,\text{sinc}(k)^2 \, e^{ikx-\frac{k^2}{1+k^2} t}
\end{equation}
yields an initial triangular density profile supported on $[-2,2]$ (as in Fig.~\ref{fig:NotReallyAcausal}), which evolves according to \eqref{nonlocal} and therefore appears to propagate superluminally (see Fig.~\ref{fig:FakingSuperlum}).

However, there is no violation of causality. The apparent propagation is an artifact of focusing only on the density. Once the full energy distribution is considered, the $\psi$ wavepacket is seen to have unbounded support already at $t=0$ (much like $\pi$ in the Klein--Gordon example, see Fig.~\ref{fig:NotReallyAcausal}). Indeed, the general solution \eqref{psiPt} shows that taking a snapshot of the $\psi$ profile at $t=a$ and $\varepsilon=b$ is equivalent to taking a snapshot at $t=0$ and $\varepsilon=a+b$. Hence, all future values of $\psi$ at a given $x$ are already encoded locally at $t=0$ in the higher-energy degrees of freedom. This is precisely the ``prior agreement'' of the stadium spectators.

\vspace{-0.2cm}
\subsection{Individual dispersion relations cannot distinguish true from fake propagation}
\vspace{-0.3cm}

At this point, the reader may feel that something is off. While model \eqref{model} does reproduce \eqref{nonlocal} for suitable initial data, there must be a fundamental difference between a system that genuinely possesses $\omega=-ik^2/(1+k^2)$ as its hydrodynamic mode, and one like ours, which merely ``fakes'' this behavior through initialization and could equally well reproduce any other dispersion relation.

From a physical perspective, such a difference indeed exists. In systems with a single hydrodynamic mode and a gapped non-hydrodynamic spectrum, the hydrodynamic mode is the longest-lived excitation and governs the late-time dynamics \cite{Dudynski1989,Geroch1995,LindblomRelaxation1996,Glorioso2018}. By contrast, the spectrum of \eqref{model} is gapless, and there is no separation of scales between hydrodynamic and non-hydrodynamic components \cite{GavassinoGapless:2024rck}.
However, if one focuses only on the analytic form of $\omega(k)$, without additional microscopic information, it is impossible to distinguish between these two scenarios. This explains why attempts to derive bounds on $\omega(k)$ based solely on its analytic form (including the attempt by \cite{HellerBounds2022ejw,HellerHydrohedron2023jtd} to derive bounds solely in terms of the convergence radius $\mathcal{R}$) are not generally valid: the analytic form of $\omega(k)$ alone does not inform us about the underlying propagation physics. Indeed, there is no fundamental mathematical distinction between the ``fake'' superluminal propagation shown in Fig.~\ref{fig:NotReallyAcausal} and that of Fig.~\ref{fig:FakingSuperlum}, as they are both wavepackets in a causal theory, fine-tuned to isolate a single dispersion branch.

\vspace{-0.4cm}
\subsection{Violating the hydrohedron bounds}
\vspace{-0.3cm}

The causal and stable model \eqref{model} produces explicit violations of the
hydrohedron inequalities when these are formulated solely in terms of the radius of convergence $\mathcal{R}$. In fact, each dispersion relation $\omega=f(k)$ is a set of correlators singularities fulfilling \eqref{StabilityBound} within their existence domain, so the theorems of \cite{HellerBounds2022ejw} should apply. However, extracting pure diffusion, $\omega=-ik^2$ (with $\mathfrak{D}=1$), yields $\mathfrak{D}\mathcal{R}=+\infty$, violating the bound $\mathfrak{D}\mathcal{R}\leq \frac{16}{3\pi}$ (i.e. Theorem 4). Examples with finite radius of convergence can also be constructed. For instance,
\begin{equation}\label{violate1}
\omega =\dfrac{-i(32 k)^2}{(3\pi)^2+(32 k)^2}
\qquad \Longrightarrow \qquad
\mathfrak{D}\mathcal{R}=\frac{32}{3\pi} \, .
\end{equation}
Remarkably, one can also violate the luminal bound on the sound speed. Carving out from \eqref{model} the dispersion relation
\begin{equation}\label{violate2}
\omega=2k-ik^2 \, ,
\end{equation}
we can mimic a viscous sound mode propagating at speed $2c$ (violating Theorem 3).

The reason these bounds fail here is the same as in the Fokker--Planck case \cite{GavassinoDiffusionCompatible:2026tvy,GavassinoFokkerPLanck:2026zsz}. Model \eqref{model} only admits modes with $\mathfrak{Im}\,\omega\leq 0$ (with $\mathfrak{Im}\,\omega=0$  included by continuity), even at complex $k$, so embeddings of, e.g., \eqref{violate1} and \eqref{violate2} are valid only for real $k$. For complex $k$, hydrodynamic modes generically violate $\mathfrak{Im}\,\omega\leq 0$ at small $k$ (at leading order, $\omega\sim k^n=|k|^n e^{in\theta}$, whose imaginary part is positive for some $\theta$), and therefore cannot be analytically continued within model \eqref{model}. The analytic continuation argument of \cite{HellerBounds2022ejw,HellerHydrohedron2023jtd} therefore does not apply.

We remark that the inability of the model to reproduce complex-$k$ solutions does not signal an incomplete embedding. Reproducing all real-$k$ solutions suffices to reconstruct, via Fourier transform, essentially any solution bounded at spatial infinity. Modes with complex $k$ grow exponentially and mainly serve to define initial-value problems for wavepackets in boosted frames, which need not exist in general (see \cite{GavassinoBoostedDiffusion:2026fff} for a detailed discussion).

\vspace{-0.2cm}
\section{Conclusions}
\vspace{-0.2cm}

We constructed a causal and covariantly-stable kinetic model whose spectrum, at real wavenumbers, contains every dispersion relation compatible with stability in the rest frame. This provides an explicit counterexample to the idea that microscopic causality can constrain the analytic form of a single dispersion relation $\omega(k)$ with no additional microscopic information. In fact, an apparently acausal mode can always arise (at real $k$) as a ``stadium wave'' of a larger causal system. In such cases, superluminal propagation reflects a particular organization of the microscopic degrees of freedom at the initial time, rather than genuine signal transmission. This highlights that causality statements are only meaningful once the set of observables defining complete local information (and hence what constitutes a ``signal'') is specified \cite{GavassinoDisperisons2023mad}. Without this specification, apparent violations of causality are ambiguous.

Our results also show that all the hydrohedron bounds (apart from those related to rest-frame stability) can cease to be applicable if the analytic continuation of a mode extends beyond the domain where a given dispersion relation corresponds to genuine modes of the microscopic theory. Nevertheless, the underlying idea of \cite{HellerBounds2022ejw} can be restored if we replace the radius of convergence $\mathcal{R}$ with a more general ``radius of validity'' $\mathcal{R}_v\leq \mathcal{R}$, which bounds the maximal disk in the complex $k$-plane where the dispersion relation admits a realization within the theory. In our model, $\mathcal{R}_v=0$, so the hydrohedron bounds are not informative; in relativistic Fokker--Planck kinetic theory, we have $\mathcal{R}_v=1/(2\mathfrak{D})$, so $\mathfrak{D}\mathcal{R}_v=\frac{1}{2}\leq \frac{16}{3\pi}$, in agreement with the bounds; in strongly coupled holographic quantum field theories, $\mathcal{R}_v$ usually coincides with $\mathcal{R}$. It is this validity radius $\mathcal{R}_v$, rather than analyticity alone, that constrains transport.

\newpage
\section*{Acknowledgements}

I thank J. Noronha, A. Serantes, and B. Withers for reading the manuscript and providing useful feedback.
This work is supported by a MERAC Foundation prize grant,  an Isaac Newton Trust Grant, and funding from the Cambridge Centre for Theoretical Cosmology.

\appendix

\section{The modes as Green-function singularities}\label{AppAAA}

In kinetic theory, retarded Green functions are constructed by perturbing the system in the origin with some distribution $A(\varepsilon)\in L^2[0,+\infty)$, yielding in our case
\begin{equation}
\psi(t,x,\varepsilon)=\Theta(t) \delta(x) A(t+\varepsilon) \, ,
\end{equation}
and then by measuring the observable $\langle B,\psi\rangle$ at some event $(t,x)$, with $B(\varepsilon)\in L^2[0,+\infty)$ and $\langle *,*\rangle$ the $L^2$ inner product. The result is the following distribution:
\begin{equation}
G_{AB}(t,x)=\Theta(t)\delta(x)\int_0^\infty B^*(\varepsilon)A(\varepsilon+t) d\varepsilon \, ,
\end{equation}
which is causal and tempered.
If we specialize to the case where $A(\varepsilon)=e^{-a\varepsilon}$ and $B(\varepsilon)=e^{-b\varepsilon}$ (with $\mathfrak{Re}a,\mathfrak{Re}b>0$), and take the Fourier-Laplace transform, we obtain
\begin{equation}
G_{AB}(\omega,k)=\int  e^{i\omega t-ikx}G_{AB}(t,x) dt dx = \dfrac{1}{(a+b^*)(a-i\omega)} \, .
\end{equation}
We see that there is a singularity at $\omega=-ia$, for which $\mathfrak{Im}\omega=-\mathfrak{Re}a< 0$. By varying $A$, one finds that for any $\omega$ in the open lower half plane, there is a corresponding Green function $G_{AB}$ that is singular there. Hence, the operator $G(\omega,k)=-(\partial_\varepsilon+i\omega)^{-1}$, whose matrix elements are $G_{AB}$, is singular on the whole lower half plane.
This is precisely the region populated by the modes \eqref{magic}.
Therefore, those modes are not merely artifacts of a particular
choice of initial data, but correspond to genuine singularities of microscopic Green functions. The role of the special initial data
used in Section \ref{stadiumwave} is only to construct effective hydrodynamic PDEs, using this pre-existing microscopic spectrum.

\bibliography{Biblio}

@article{Sommerfeld1907,
  author = {A. Sommerfeld},
  title = {\"Uber die Fortpflanzung des Lichtes in dispergierenden Medien},
  journal = {Annalen der Physik},
  volume = {327},
  number = {2},
  pages = {177--202},
  year = {1907},
  doi = {10.1002/andp.19073270202}
}

@article{BludmanRuderman1968,
  author = {S. A. Bludman and M. A. Ruderman},
  title = {Possibility of the Speed of Sound Exceeding the Speed of Light in Ultradense Matter},
  journal = {Physical Review},
  volume = {170},
  number = {5},
  pages = {1176--1184},
  year = {1968},
  doi = {10.1103/PhysRev.170.1176}
}

@article{Fox1970,
  author = {R. Fox and C. G. Kuper and S. G. Lipson},
  title = {Faster-than-light group velocities and causality violation},
  journal = {Proceedings of the Royal Society of London A},
  volume = {316},
  number = {1527},
  pages = {515--524},
  year = {1970},
  doi = {10.1098/rspa.1970.0093}
}

@article{Susskind1969,
  title = {Superluminal Behavior, Causality, and Instability},
  author = {Aharonov, Y. and Komar, A. and Susskind, L.},
  journal = {Phys. Rev.},
  volume = {182},
  issue = {5},
  pages = {1400--1403},
  numpages = {0},
  year = {1969},
  month = {Jun},
  publisher = {American Physical Society},
  doi = {10.1103/PhysRev.182.1400},
  url = {https://link.aps.org/doi/10.1103/PhysRev.182.1400}
}

@book{Brillouin1960,
  author = {L. Brillouin},
  title = {Wave Propagation and Group Velocity},
  year = {1960},
  publisher = {Academic Press}
}

@article{DudynskiCausality1985,
  title = {Causality of the linearized relativistic Boltzmann equationp},
  author = {Dudynski, Marek and Ekiel-Jez\ifmmode \grave{}\else \`{}\fi{}ewska, Maria L.},
  journal = {Phys. Rev. Lett.},
  volume = {55},
  issue = {26},
  pages = {2831--2834},
  numpages = {0},
  year = {1985},
  month = {Dec},
  publisher = {American Physical Society},
  doi = {10.1103/PhysRevLett.55.2831},
  url = {https://link.aps.org/doi/10.1103/PhysRevLett.55.2831}
}

@article{DisconziReview:2023rtt,
    author = "Disconzi, Marcelo M.",
    title = "{Recent developments in mathematical aspects of relativistic fluids}",
    eprint = "2308.09844",
    archivePrefix = "arXiv",
    primaryClass = "math.AP",
    doi = "10.1007/s41114-024-00052-x",
    journal = "Living Rev. Rel.",
    volume = "27",
    number = "1",
    pages = "6",
    year = "2024"
}

@article{Wang:2023csj,
    author = "Wang, Dong-Lin and Pu, Shi",
    title = "{Stability and causality criteria in linear mode analysis: Stability means causality}",
    eprint = "2309.11708",
    archivePrefix = "arXiv",
    primaryClass = "hep-th",
    doi = "10.1103/PhysRevD.109.L031504",
    journal = "Phys. Rev. D",
    volume = "109",
    number = "3",
    pages = "L031504",
    year = "2024"
}

@article{HoultClassicalDispersion:2023clg,
    author = "Hoult, Raphael E. and Kovtun, Pavel",
    title = "{Causality and classical dispersion relations}",
    eprint = "2309.11703",
    archivePrefix = "arXiv",
    primaryClass = "hep-th",
    doi = "10.1103/PhysRevD.109.046018",
    journal = "Phys. Rev. D",
    volume = "109",
    number = "4",
    pages = "046018",
    year = "2024"
}

@article{Aifantis1980,
  author    = {E. C. Aifantis},
  title     = {On the problem of diffusion in solids},
  journal   = {Acta Mechanica},
  year      = {1980},
  volume    = {37},
  number    = {3-4},
  pages     = {265--296},
  doi       = {10.1007/BF01202949},
  url       = {https://doi.org/10.1007/BF01202949},
  publisher = {Springer}
}

@article{Ting197423,
title = {A cooling process according to two-temperature theory of heat conduction},
journal = {Journal of Mathematical Analysis and Applications},
volume = {45},
number = {1},
pages = {23-31},
year = {1974},
issn = {0022-247X},
doi = {https://doi.org/10.1016/0022-247X(74)90116-4},
url = {https://www.sciencedirect.com/science/article/pii/0022247X74901164},
author = {Tsuan Wu Ting}
}

@article{Barenblatt1963OnCB,
  title={On certain boundary-value problems for the equations of seepage of a liquid in fissured rocks},
  author={G. I. Barenblatt},
  journal={Journal of Applied Mathematics and Mechanics},
  year={1963},
  volume={27},
  pages={513-518},
  url={https://api.semanticscholar.org/CorpusID:119763685}
}

@article{Barenblatt1960BasicCI,
  title={Basic concepts in the theory of seepage of homogeneous liquids in fissured rocks [strata]},
  author={G. I. Barenblatt and I P Zheltov and I. N. Kochina},
  journal={Journal of Applied Mathematics and Mechanics},
  year={1960},
  volume={24},
  pages={1286-1303},
  url={https://api.semanticscholar.org/CorpusID:120893886}
}

@article{Dudynski1989,
  author       = {Dudy{\'n}ski, Marek},
  title        = {On the Linearized Relativistic Boltzmann Equation. II. Existence of Hydrodynamics},
  journal      = {Journal of Statistical Physics},
  volume       = {57},
  number       = {1-2},
  pages        = {199--245},
  year         = {1989},
  publisher    = {Springer},
  url          = {https://link.springer.com/article/10.1007/BF01023641#citeas},
  doi          = {10.1007/BF01023641}
}

@article{Hegerfeldt1974,
  author       = {Hegerfeldt, Gerhard C.},
  title        = {Remark on Causality and Particle Localization},
  journal      = {Physical Review D},
  volume       = {10},
  number       = {10},
  pages        = {3320--3321},
  year         = {1974},
  doi          = {10.1103/PhysRevD.10.3320}
}

@article{GavassinoFokkerPLanck:2026zsz,
    author = "Gavassino, Lorenzo",
    title = "{The quasi-normal modes of relativistic Fokker-Planck kinetic theory}",
    eprint = "2601.19474",
    archivePrefix = "arXiv",
    primaryClass = "nucl-th",
    month = "1",
    year = "2026"
}

@article{GavassinoBoostedDiffusion:2026fff,
    author = "Gavassino, Lorenzo",
    title = "{Lorentz-boosted diffusion: initial value formulation and exact solutions}",
    eprint = "2602.21254",
    archivePrefix = "arXiv",
    primaryClass = "math-ph",
    month = "2",
    year = "2026"
}

@article{GavassinoDiffusionCompatible:2026tvy,
    author = "Gavassino, Lorenzo",
    title = "{The diffusion equation is compatible with special relativity}",
    eprint = "2601.19464",
    archivePrefix = "arXiv",
    primaryClass = "gr-qc",
    month = "1",
    year = "2026"
}

@article{GavassinoBounds2023myj,
    author = "Gavassino, L.",
    title = "{Bounds on transport from hydrodynamic stability}",
    eprint = "2301.06651",
    archivePrefix = "arXiv",
    primaryClass = "hep-th",
    doi = "10.1016/j.physletb.2023.137854",
    journal = "Phys. Lett. B",
    volume = "840",
    pages = "137854",
    year = "2023"
}

@article{HellerHydrohedron2023jtd,
    author = "Heller, Michal P. and Serantes, Alexandre and Spali\'nski, Micha\l{} and Withers, Benjamin",
    title = "{The space of transport coefficients allowed by causality}",
    eprint = "2305.07703",
    archivePrefix = "arXiv",
    primaryClass = "hep-th",
    doi = "10.1038/s41567-024-02635-5",
    journal = "Nature Phys.",
    volume = "20",
    number = "12",
    pages = "1948--1954",
    year = "2024"
}

@ARTICLE{Pu2010,
       author = {{Pu}, Shi and {Koide}, Tomoi and {Rischke}, Dirk H.},
        title = "{Does stability of relativistic dissipative fluid dynamics imply causality?}",
      journal = {\prd},
         year = 2010,
        month = jun,
       volume = {81},
       number = {11},
          eid = {114039},
        pages = {114039},
          doi = {10.1103/PhysRevD.81.114039},
archivePrefix = {arXiv},
       eprint = {0907.3906},
 primaryClass = {hep-ph},
       adsurl = {https://ui.adsabs.harvard.edu/abs/2010PhRvD..81k4039P}
}

@ARTICLE{GavassinoSuperlum2021,
    author = "Gavassino, Lorenzo",
    title = "{Can We Make Sense of Dissipation without Causality?}",
    eprint = "2111.05254",
    archivePrefix = "arXiv",
    primaryClass = "gr-qc",
    doi = "10.1103/PhysRevX.12.041001",
    journal = "Phys. Rev. X",
    volume = "12",
    number = "4",
    pages = "041001",
    year = "2022"
}

@misc{Brants2025SavingCausality,
  author       = {Robbe Brants},
  title        = {Cutting Through the Frequency Plane to Save Causality},
  year         = {2025},
  note         = {Conference presentation at Quantum Dynamics at Ghent 2025},
  url          = {https://indico.global/event/13082/contributions/115394/attachments/56090/107736/saving%20causality.pdf}
}

@ARTICLE{Geroch1995,
       author = {{Geroch}, Robert},
        title = "{Relativistic theories of dissipative fluids}",
      journal = {Journal of Mathematical Physics},
         year = 1995,
        month = aug,
       volume = {36},
       number = {8},
        pages = {4226-4241},
          doi = {10.1063/1.530958},
       adsurl = {https://ui.adsabs.harvard.edu/abs/1995JMP....36.4226G}
}

@ARTICLE{Krotscheck1978,
       author = {{Krotscheck}, Eckhard and {Kundt}, Wolfgang},
        title = "{Causality criteria}",
      journal = {Communications in Mathematical Physics},
         year = 1978,
        month = jun,
       volume = {60},
       number = {2},
        pages = {171-180},
          doi = {10.1007/BF01609447},
       adsurl = {https://ui.adsabs.harvard.edu/abs/1978CMaPh..60..171K}
}

@ARTICLE{GavassinoGibbs2021,
      author = {{Gavassino}, L.},
        title = "{Applying the Gibbs stability criterion to relativistic hydrodynamics}",
      journal = {Classical and Quantum Gravity},
         year = 2021,
        month = nov,
       volume = {38},
       number = {21},
          eid = {21LT02},
        pages = {21LT02},
          doi = {10.1088/1361-6382/ac2b0e},
archivePrefix = {arXiv},
       eprint = {2104.09142},
 primaryClass = {gr-qc},
       adsurl = {https://ui.adsabs.harvard.edu/abs/2021CQGra..38uLT02G}
}

@ARTICLE{GavassinoCausality2021,
    author = "Gavassino, Lorenzo and Antonelli, Marco and Haskell, Brynmor",
    title = "{Thermodynamic Stability Implies Causality}",
    eprint = "2105.14621",
    archivePrefix = "arXiv",
    primaryClass = "gr-qc",
    doi = "10.1103/PhysRevLett.128.010606",
    journal = "Phys. Rev. Lett.",
    volume = "128",
    number = "1",
    pages = "010606",
    year = "2022"
}

@book{Groot1980RelativisticKT,
  title={Relativistic kinetic theory: principles and applications},
  author={S. R. de Groot and W. Astrid van Leeuwen and Christianus G. van Weert},
  year={1980}
}

@ARTICLE{LindblomRelaxation1996,
       author = {{Lindblom}, Lee},
        title = "{The Relaxation Effect in Dissipative Relativistic Fluid Theories}",
      journal = {Annals of Physics},
         year = 1996,
        month = apr,
       volume = {247},
       number = {1},
        pages = {1-18},
          doi = {10.1006/aphy.1996.0036},
archivePrefix = {arXiv},
       eprint = {gr-qc/9508058},
 primaryClass = {gr-qc},
       adsurl = {https://ui.adsabs.harvard.edu/abs/1996AnPhy.247....1L}
}

@article{Bemfica2019_conformal1,
  title = {Nonlinear causality of general first-order relativistic viscous hydrodynamics},
  author = {Bemfica, F\'abio S. and Disconzi, Marcelo M. and Noronha, Jorge},
  journal = {Phys. Rev. D},
  volume = {100},
  issue = {10},
  pages = {104020},
  numpages = {13},
  year = {2019},
  month = {Nov},
  publisher = {American Physical Society},
  doi = {10.1103/PhysRevD.100.104020},
  url = {https://link.aps.org/doi/10.1103/PhysRevD.100.104020}
}

@article{GavassinoLyapunov_2020,
   title={When the entropy has no maximum: A new perspective on the instability of the first-order theories of dissipation},
   volume={102},
   ISSN={2470-0029},
   url={http://dx.doi.org/10.1103/PhysRevD.102.043018},
   DOI={10.1103/physrevd.102.043018},
   number={4},
   journal={Physical Review D},
   publisher={American Physical Society (APS)},
   author={Gavassino, L. and Antonelli, M. and Haskell, B.},
   year={2020},
   month={Aug}
}

@article{Hishcock1983,
title = "Stability and causality in dissipative relativistic fluids",
journal = "Annals of Physics",
volume = "151",
number = "2",
pages = "466 - 496",
year = "1983",
issn = "0003-4916",
doi = "https://doi.org/10.1016/0003-4916(83)90288-9",
url = "http://www.sciencedirect.com/science/article/pii/0003491683902889",
author = "William A. Hiscock and Lee Lindblom"
}

@article{GavassinoGapless:2024rck,
    author = "Gavassino, Lorenzo",
    title = "{Gapless nonhydrodynamic modes in relativistic kinetic theory}",
    eprint = "2404.12327",
    archivePrefix = "arXiv",
    primaryClass = "nucl-th",
    doi = "10.1103/PhysRevResearch.6.L042043",
    journal = "Phys. Rev. Res.",
    volume = "6",
    number = "4",
    pages = "L042043",
    year = "2024"
}

@article{GavassinoDisperisons2023mad,
    author = "Gavassino, Lorenzo and Disconzi, Marcelo M. and Noronha, Jorge",
    title = "{Dispersion Relations Alone Cannot Guarantee Causality}",
    eprint = "2307.05987",
    archivePrefix = "arXiv",
    primaryClass = "hep-th",
    doi = "10.1103/PhysRevLett.132.162301",
    journal = "Phys. Rev. Lett.",
    volume = "132",
    number = "16",
    pages = "162301",
    year = "2024"
}

@ARTICLE{Glorioso2018,
       author = {{Glorioso}, Paolo and {Liu}, Hong},
        title = "{Lectures on non-equilibrium effective field theories and fluctuating hydrodynamics}",
      journal = {arXiv e-prints},
         year = 2018,
        month = may,
          eid = {arXiv:1805.09331},
        pages = {arXiv:1805.09331},
archivePrefix = {arXiv},
       eprint = {1805.09331},
 primaryClass = {hep-th},
       adsurl = {https://ui.adsabs.harvard.edu/abs/2018arXiv180509331G}
}

@article{HellerBounds2022ejw,
    author = "Heller, Michal P. and Serantes, Alexandre and Spali{\'n}ski, Micha{\l} and Withers, Benjamin",
    title = "{Rigorous Bounds on Transport from Causality}",
    eprint = "2212.07434",
    archivePrefix = "arXiv",
    primaryClass = "hep-th",
    doi = "10.1103/PhysRevLett.130.261601",
    journal = "Phys. Rev. Lett.",
    volume = "130",
    number = "26",
    pages = "261601",
    year = "2023"
}

@book{lasalle1961stability,
  title={Stability by Liapunov's Direct Method: With Applications},
  author={LaSalle, J.P. and Lefschetz, S.},
  lccn={61018296},
  series={Mathematics in science andengineering, v.4},
  url={https://books.google.pl/books?id=fAZRAAAAMAAJ},
  year={1961},
  publisher={Academic Press}
}

\appendix

\label{lastpage}
\end{document}